\documentclass[aps,onecolumn]{revtex4}
\usepackage{amsfonts}
\usepackage{amsmath}
\usepackage{amssymb}
\usepackage{version}
\usepackage{graphicx}

\includeversion{New_connection}
\excludeversion{Old_connection}

\begin{document}

\preprint{}
\title{Switching Current Distributions in Josephson Junctions at Low
Temperatures Resulting From Noise Enhanced Thermal Activation }
\author{James A. Blackburn}
\affiliation{Department of Physics and Computer Science\\
Wilfrid Laurier University\\
Waterloo, Ontario, Canada}

\begin{abstract}
Experiments on the distributions of switching currents in Josephson
junctions are sensitive probes of the mechanism by which a junction changes
abruptly to a finite voltage state. At low temperatures data exhibit smooth
and gradual deviations from the expectations of the classical theory of
thermal activation over the barrier in the tilted washboard potential. \ In
this paper it is shown that if a very small proportion of the noise energy
entering the apparatus at room temperature survives filtering\ and reaches
the sample, it can enhance the escape rate sufficiently to replicate
experimental observations of the temperature dependence of the switching
bias. This conjecture is successfully tested against published experimental
data.
\end{abstract}

\maketitle

\section{Introduction}

When the bias current applied to a Josephson junction is gradually
increased, the junction will eventually switch to a non-zero voltage state.
For repeated trials carried out at the same temperature, each commencing at
zero bias current, this switching occurs at slightly different values of
bias current. \ Accumulated escape data form a switching current
distribution (SCD) whose peak defines the most probable value for the escape
bias. For temperatures above about $100mK$, these characteristics are well
described by the classical theory of thermally activated escape out of the
washboard potential, as discussed in the next section.

When dilution refrigerators became available, experiments were possible down
to base temperatures as small as a few millikelvin. Leggett \cite{Leggett}
had predicted that below a \textquotedblleft crossover
temperature\textquotedblright\ a Josephson junction would enter a
macroscopic quantum state.\ Voss and Webb \cite{VossWebb} were the first to
claim confirmation of this conjecture with evidence based on observations
reaching $5mK$ and interpreted with the hypothesis of macroscopic quantum
tunneling (MQT) as the new mechanism for escape from the well.

But as has already been pointed out \cite{BCJJAP},\cite{Blackburn} the data
in swept bias experiments \cite{VossWebb},\cite{Yu},\cite{Oelsner} do not,
on very close inspection, exhibit the temperature independence inherent in
MQT theory. In fact SCD escape peaks always retained some slight temperature
dependence. \ Therefore this crucial attribute of peak freezing had not been
observed, thus casting doubt on the conjecture of a crossover.

At such low temperatures, the SCD peak behavior was instead observed to
deviate smoothly from the classical prediction, and this fact requires an
explanation. A \emph{modified} classical thermal escape rate is proposed
here and is successfully tested against data from three independent
experiments.

\section{Thermal Activation}

Thermal activation from a potential well at a given temperature is governed,
in the low damping regime, by the well known escape rate due to Kramers \cite%
{Kramers}.

\begin{equation}
\Gamma =f_{J}\exp \left( -\frac{\Delta U}{k_{B}T}\right)  \label{Kramers}
\end{equation}%
where $\Delta U$ is the barrier height and, for a Josephson junction, $\
f_{J}$ is the plasma frequency $f_{J}=f_{J0}\left( 1-\eta ^{2}\right) ^{1/4}$%
, with $\eta =I/I_{C}$ denoting the normalized bias current. The barrier
height in the Josephson washboard potential is given by

\begin{equation}
\Delta U=2E_{J}\left( \sqrt{1-\eta ^{2}}-\eta \cos ^{-1}\eta \right)
\label{DU}
\end{equation}%
and $E_{J}=\hbar I_{C}/2e$ is the Josephson energy.

\begin{figure}
[h]
\begin{center}
\includegraphics[
height=2.6521in,
width=3.4326in
]
{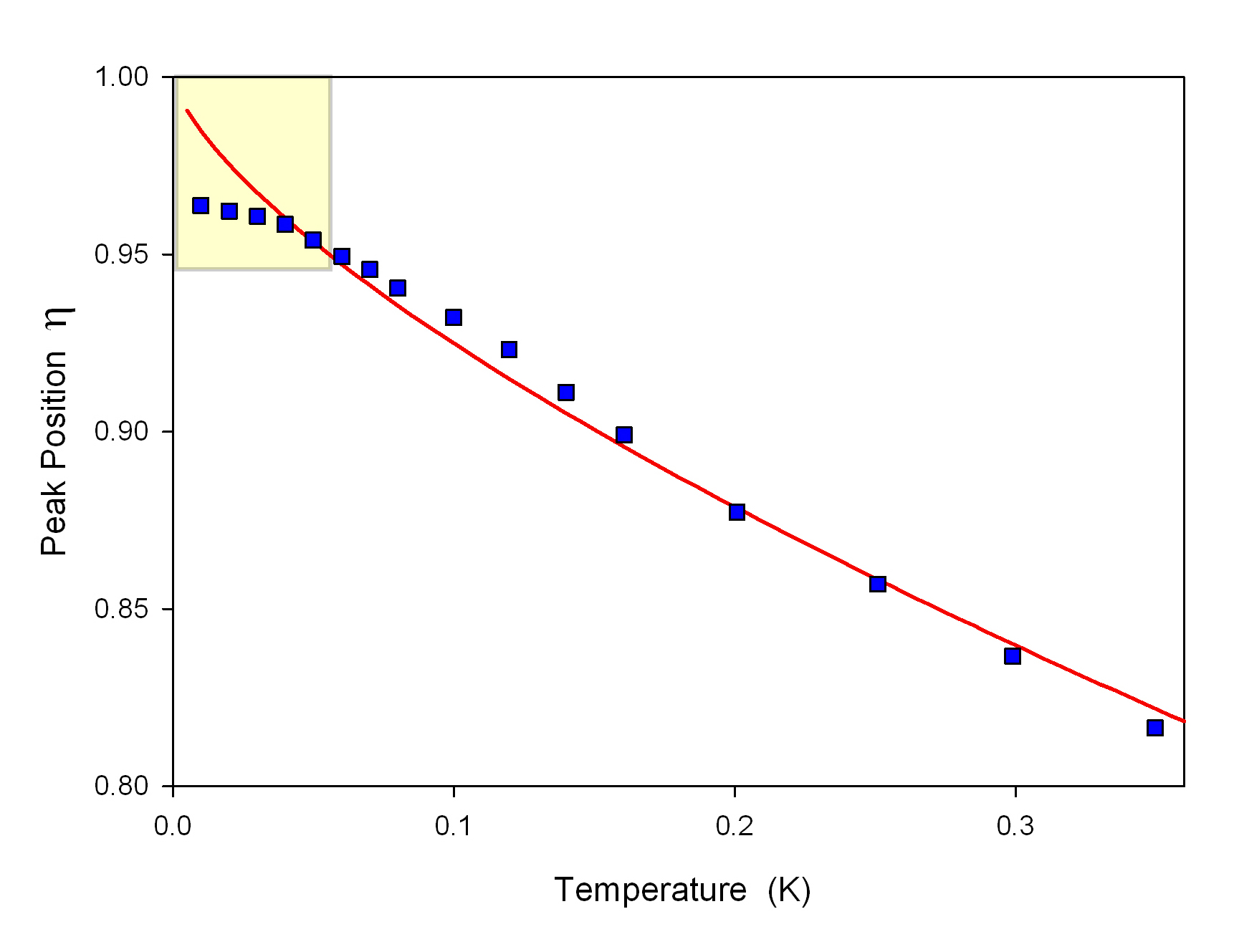}
\caption{Temperature
dependence of peak positions in switching current distributions obtained
from a simulation of the experiment of Oelsner et al. (solid line) using the
escape rate given in Eq.(1), together with \ experimental data points
(squares). The shaded area highlights the peeling away from the Kramers
thermal activation prediction that occurs at the lowest temperatures.}
\label{OelsnerSCD}
\end{center}
\end{figure}

\begin{figure}
[h]
\begin{center}
\includegraphics[
height=2.2858in,
width=5.9641in
]
{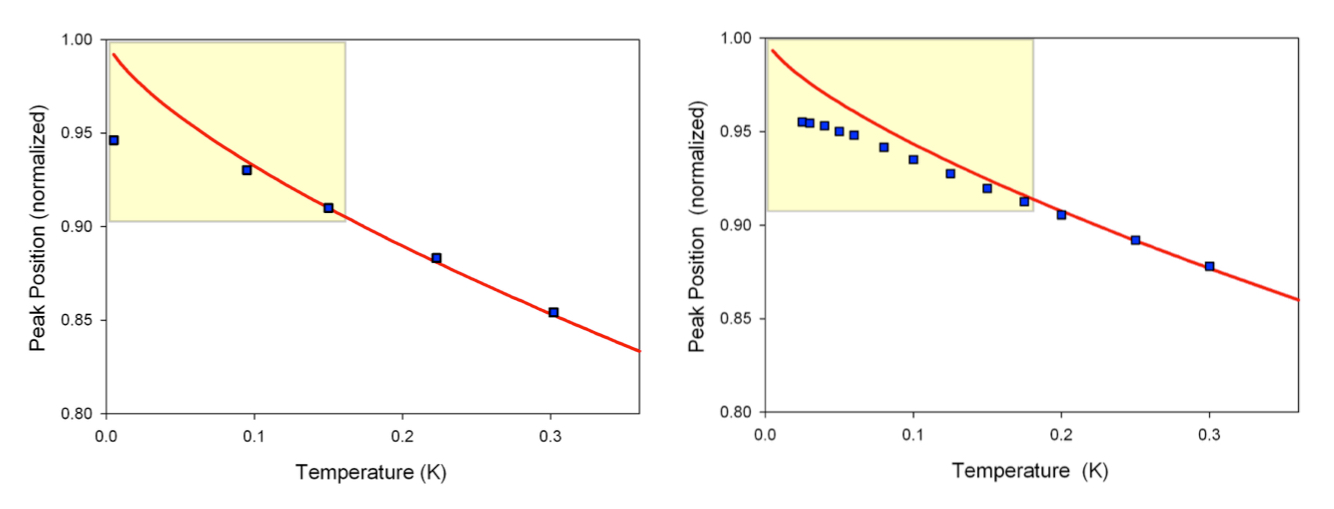}
\caption{Experimental data (squares) and
simulation results (lines) from Voss \& Webb (left panel) and Yu et al.
(right panel).}
\label{vw_yu}
\end{center}
\end{figure}

Fig.\ref{OelsnerSCD} displays the output from an algorithm-based simulation 
\cite{BCJPRB}, \cite{BCJPR} (see \cite{footnote}) carried out with the
experimental parameters of Oelsner et al. \cite{Oelsner} and the escape rate
given in Eq.(\ref{Kramers}). \ Also shown are the data points from this
experiment, digitized from Fig.1 in \cite{Oelsner} and converted to a linear
temperature scale. 
The gradual peeling away of
experimental data from the expectation based on the standard escape rate
represents a deviation from classical thermal activation at the lowest
temperatures. \ Such smooth peeling away is also observed in the experiments
of Voss and Webb \cite{VossWebb} and Yu et al.\cite{Yu} - see Fig.\ref{vw_yu}

With linear temperature scales, it is very apparent that the deviation of
experimental data from standard thermal activation characteristics is
gradual and smooth, without any sign of an abrupt crossover.

\section{Escape Rates}

\begin{figure}
[h]
\begin{center}
\includegraphics[
height=2.1288in,
width=4.1458in
]
{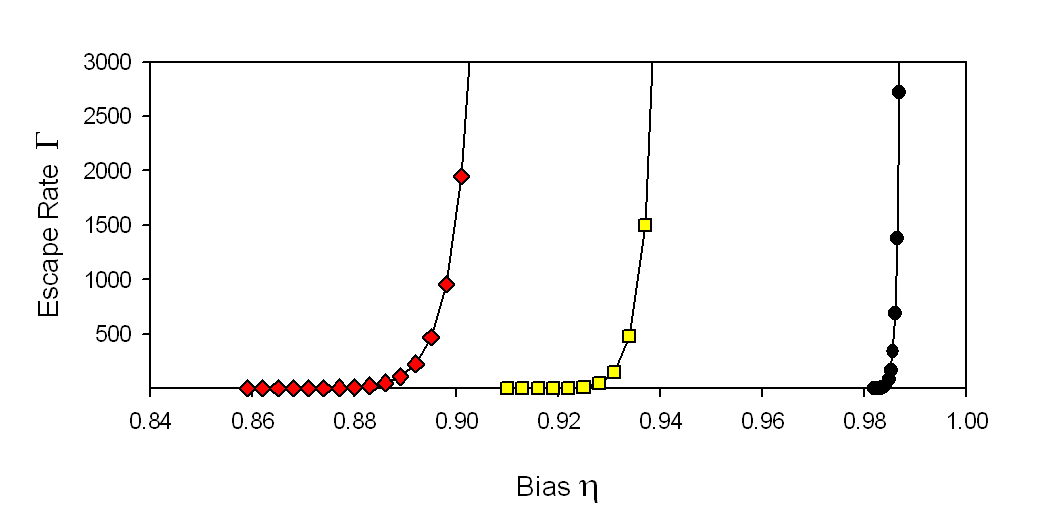}
\caption{Escape rate $\Gamma $ as a
function of normalized bias $\protect\eta $ for temperatures $T=0.200K$
(diamonds), $T=0.100K$ (squares), and $T=0.010K$ (circles). \ The junction
critical current and zero bias plasma frequency were taken from Oelsner et
al.}
\label{Gamma1}
\end{center}
\end{figure}

The escape rate $\Gamma $ is a static function that can be evaluated
directly from Eqs.(\ref{Kramers} and \ref{DU}). In Fig. \ref{Gamma1}, escape
rates are calculated with the parameters of Oelsner et al. \cite{Oelsner}. 
Each plot represents the evolution of 
$\Gamma $over the course of a bias sweep at some specified temperature. As
can be seen, the escape rate remains negligible until, in the neighbourhood
of some bias, $\Gamma $ increases very rapidly. \ This upswing is
comparatively gradual at higher temperatures and becomes more abrupt as the
temperature is lowered. \ This property is the cause of SCD peaks becoming
sharper as temperature decreases. As simulations reveal, the peak in a
switching current distrbution is located close to the bias value at this
abrupt upswing.

However, in Fig.\ref{Gamma1} the characteristic at $T=10mK$ \ has an onset
at about $\eta =0.985$ which, it turns out, is too high; the experimental
peak is closer to $\eta =0.964$. \ Escape rates extracted from numerical
solutions of the Langevin equation \cite{Langevin} at this temperature
matched the curve in Fig. \ref{Gamma1} and thus confirmed that the simple
Kramers expression is still valid even down to such a low temperature \cite%
{Chungho}.\ \ Therefore, the origin of the discrepancy must be sought
elsewhere. \ 

\section{Enhanced Thermal Activation}

\begin{figure}
[h]
\begin{center}
\includegraphics[
height=2.5589in,
width=3.4246in
]
{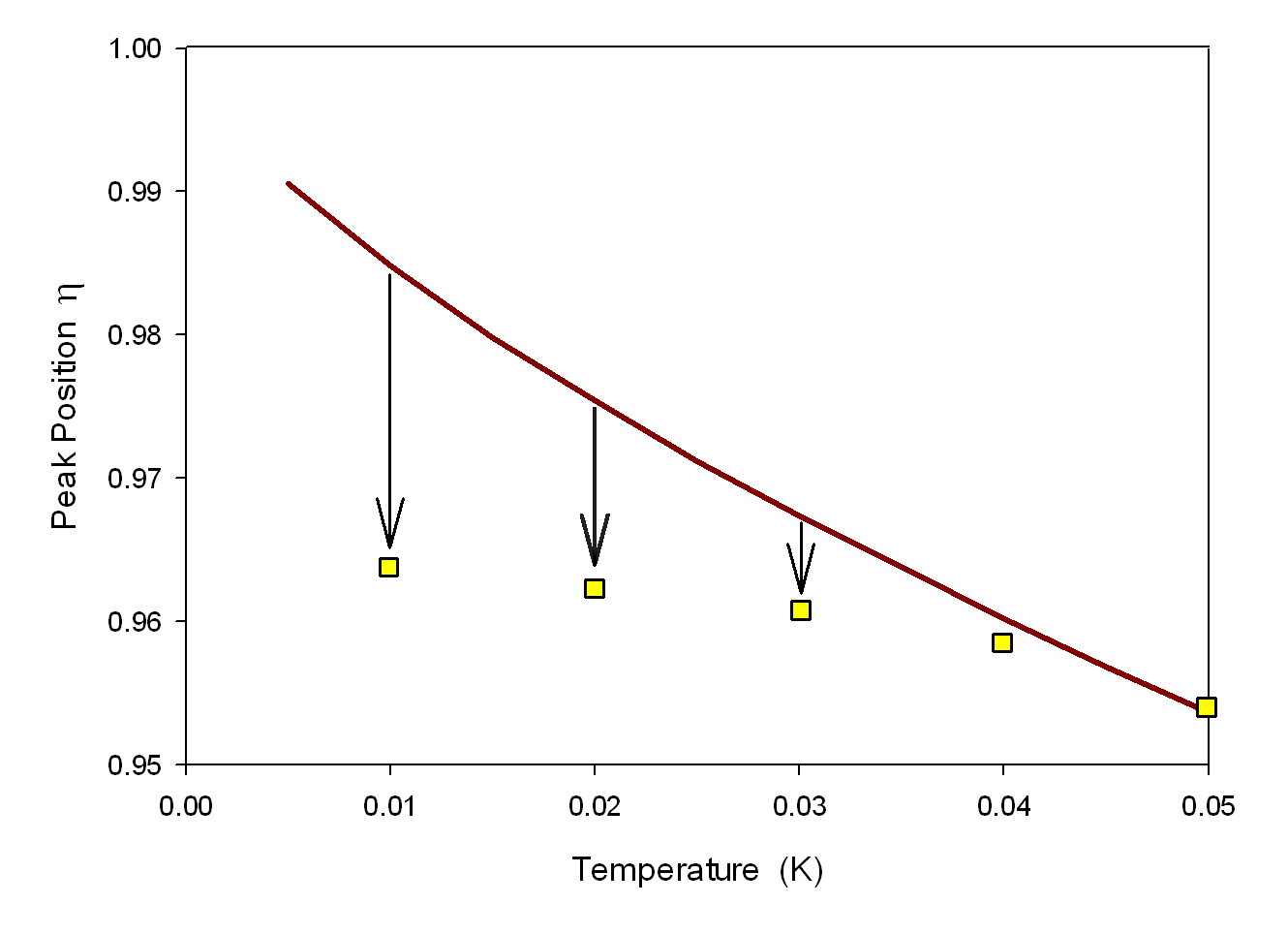}
\caption{Low temperature region of the simulation
shown in Fig.1 together with data points (squares) from the experiment of
Oelsner et al.\ The arrows indicate the required lowering of the simulated
peak positions at these temperatures.}
\label{adjustment}
\end{center}
\end{figure}

For high temperatures, simulation results closely match experimental data. \
But as indicated in the previous section, discrepancies exist at the lowest
temperatures. This is illustrated in Fig.\ref{adjustment} which is based on
the same experiments \cite{Oelsner} used for Fig. \ref{OelsnerSCD}. 

Clearly, the simulation based on the escape rate in Eq.\ref%
{Kramers} does not agree with observations at the lowest temperatures. As
indicated by the arrows in Fig.\ref{adjustment}, the simulation predicts SCD
peaks lying above the observed switching current distributions, therefore an
enhanced escape rate is implied.

The following ansatz is proposed:

\begin{equation}
\Gamma =f_{J}\exp \left( -\varepsilon \frac{\Delta U}{k_{B}T}\right)
\label{Kramers2}
\end{equation}%
where the parameter $\varepsilon $ will be $1.0$ for standard thermal
activation and $\varepsilon <1$ when there is an \textit{increase} in the
escape rate. So thermal activation would be \emph{enhanced} for $%
0<\varepsilon <1$.

To bring simulation into agreement with observation, the following sequence
was followed: select a temperature; begin with $\varepsilon =1$ and run a
simulation to find the SCD preak position $\eta _{P}$; iteratively decrease
the value of the parameter $\varepsilon $ in small steps, performing
simulated bias sweeps and stop when the simulated peak closely matches the
experimental value for that temperature; repeat for each temperature in the
experimental dataset.

\begin{figure}
[h]
\begin{center}
\includegraphics[
height=2.9483in,
width=4.0171in
]
{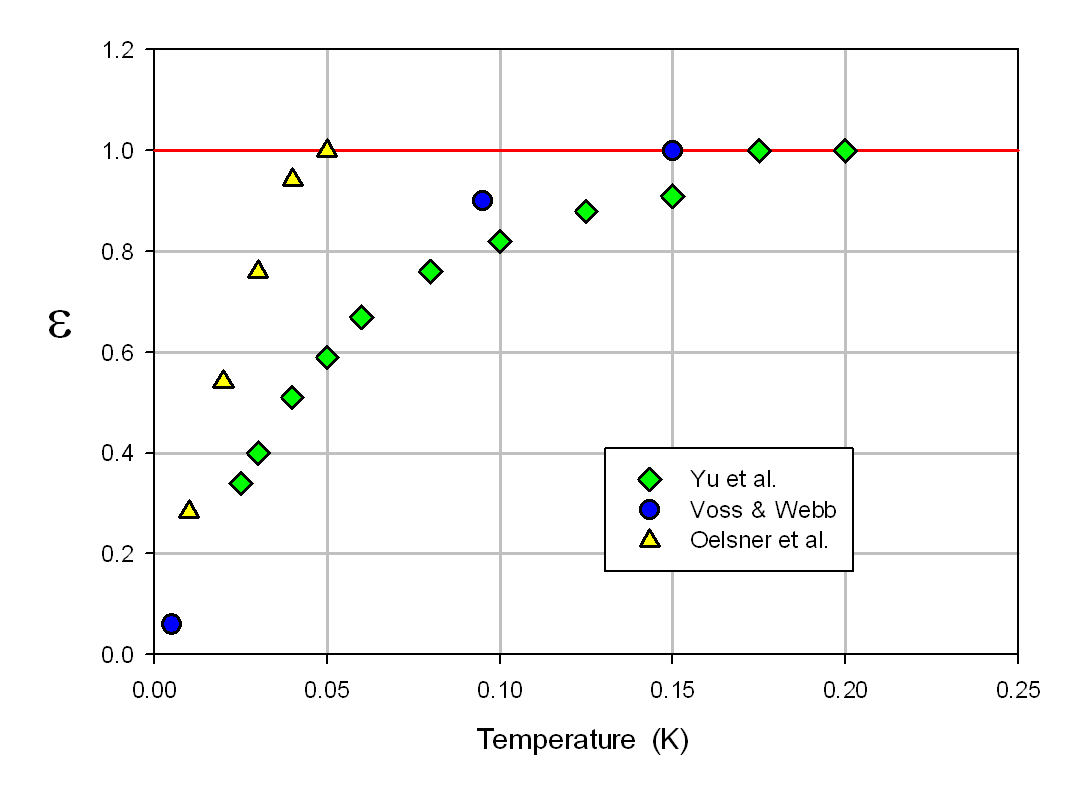}
\caption{Escape rate enhancement parameter $\protect%
\varepsilon $ as a function of bath temperature for three experiments: Yu et
al. (diamond), Voss \& Webb (circles), Oelsner et al. (triangles).}
\label{epsilon}
\end{center}
\end{figure}

\bigskip This iterative process was carried out for the experiments of Voss
and Webb \cite{VossWebb}, Yu et al. \cite{Yu}, and Oelsner et al.\cite%
{Oelsner}. \ The results are displayed in Fig.\ref{epsilon}\

The line at $\varepsilon =1.0$
indicates when simulations based on Eq.(\ref{Kramers}) will accurately match
the observations.

\begin{figure}
[h]
\begin{center}
\includegraphics[
height=2.1004in,
width=4.1458in
]
{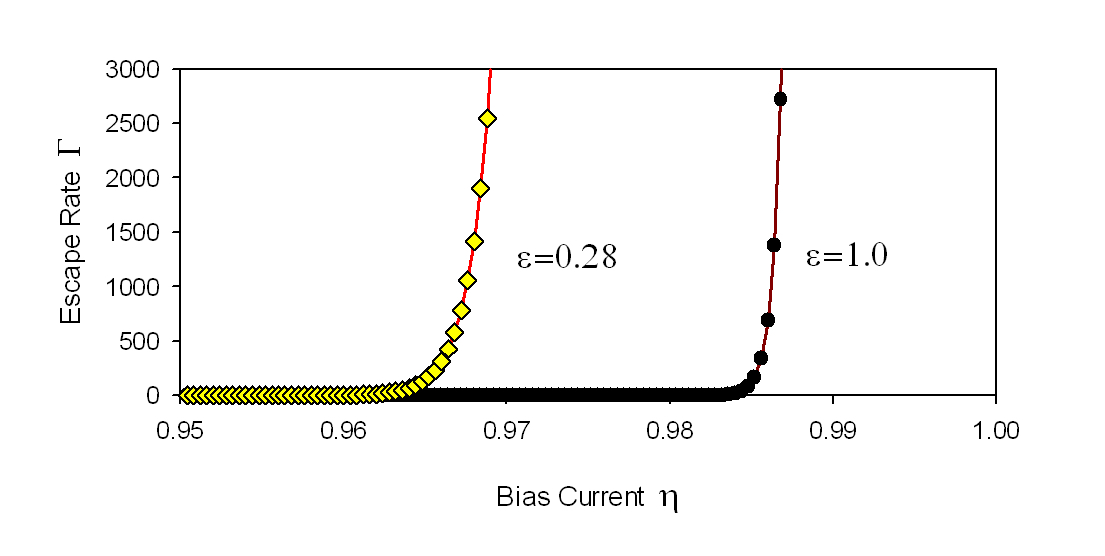}
\caption{Thermal
activation escape rates, with and without the correction factor $\protect%
\varepsilon $, for the parameters of Oelsner et al.}
\label{Gamma}
\end{center}
\end{figure}

The procedure is illustrated by the following example based on the data of
Oelsner et al.\cite{Oelsner}. As noted earlier with respect to Fig.\ref%
{Gamma1}, the $10mK$ escape rate `switched on' at too high a bias current.
From Fig. \ref{epsilon} it can be seen that at this value of $T$, $\
\varepsilon =0.28$\ is appropriate to correct this problem. Fig.\ref{Gamma}
illustrates the change from an unmodified $\Gamma $ as defined in Eq.(\ref%
{Kramers}) to a modified expression Eq.(\ref{Kramers2}) - \ the upswing
point can be seen to shift down to agree with the experimental SCD position
of $\eta _{P}=0.964$.

\begin{figure}
[h]
\begin{center}
\includegraphics[
height=2.7966in,
width=3.7688in
]
{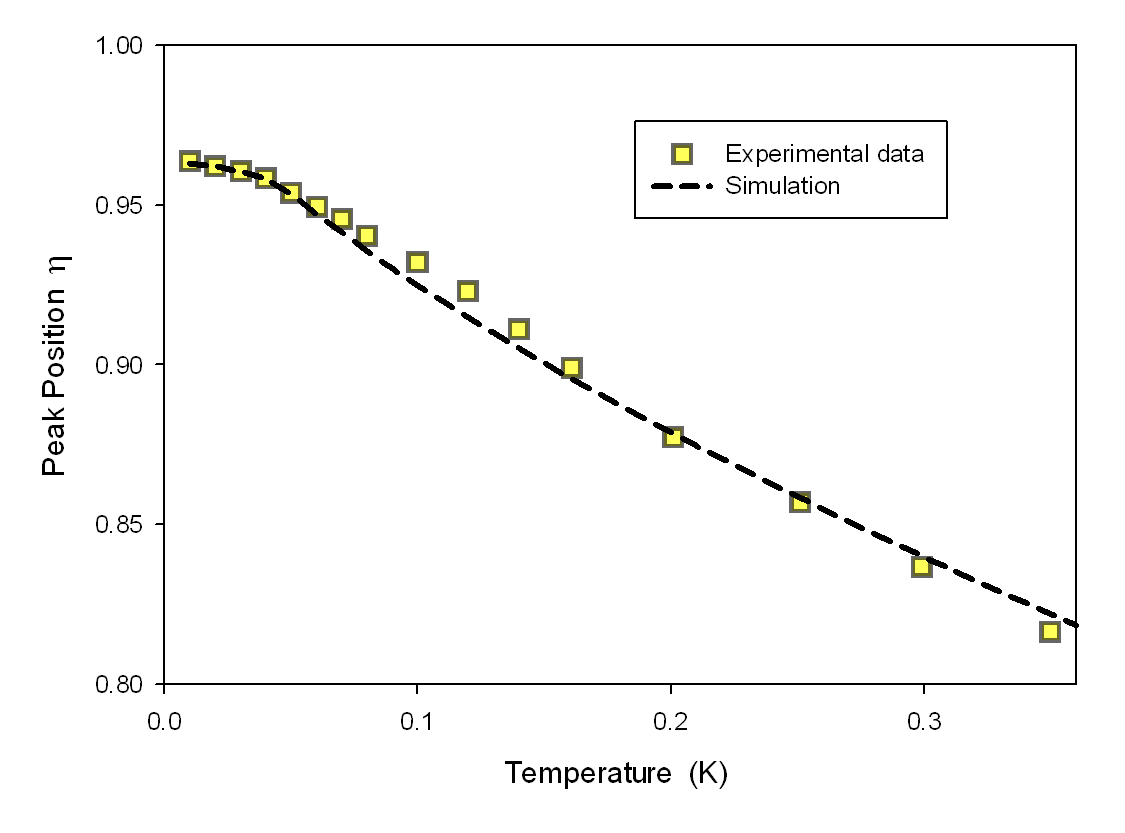}
\caption{Final
results combining a simulation that uses $\protect\varepsilon $, and
experimental data from Oelsner et al.}
\label{corrected}
\end{center}
\end{figure}

The values of $\varepsilon $ determined in Fig.\ref{epsilon} for the
experiment of Oelsner et al. \cite{Oelsner} were applied in a full
simulation covering a broader range of temperatures. \ The results are shown
in Fig.\ref{corrected}. 

As can be seen, the
simulation precisely matches the experimental data, even below $50mK$.

\section{External Noise}

No experimental system consisting of a sample chamber in a dilution
refrigerator and shielded from the room temperature world by means of
isolation stages \cite{stages} and powder filters \cite{filters} can
actually achieve a perfect noise free state. Some noise energy is bound to
reach the sample chamber. \ The following two quotes bear on this issue:

\begin{enumerate}
\item \textquotedblleft Moreover, the available concepts often provide
inadequate filtering to operate at temperatures below $10mK$%
\textquotedblright\ from \cite{filters2}

\item "A residual noise temperature $T_{N}$, which can be minimized but
never fully eliminated in this kind of experiments, can also lead to some
ambiguities at low temperatures" \ from \cite{Silvestrini}.
\end{enumerate}

\begin{figure}
[h]
\begin{center}
\includegraphics[
height=3.0539in,
width=4.1555in
]
{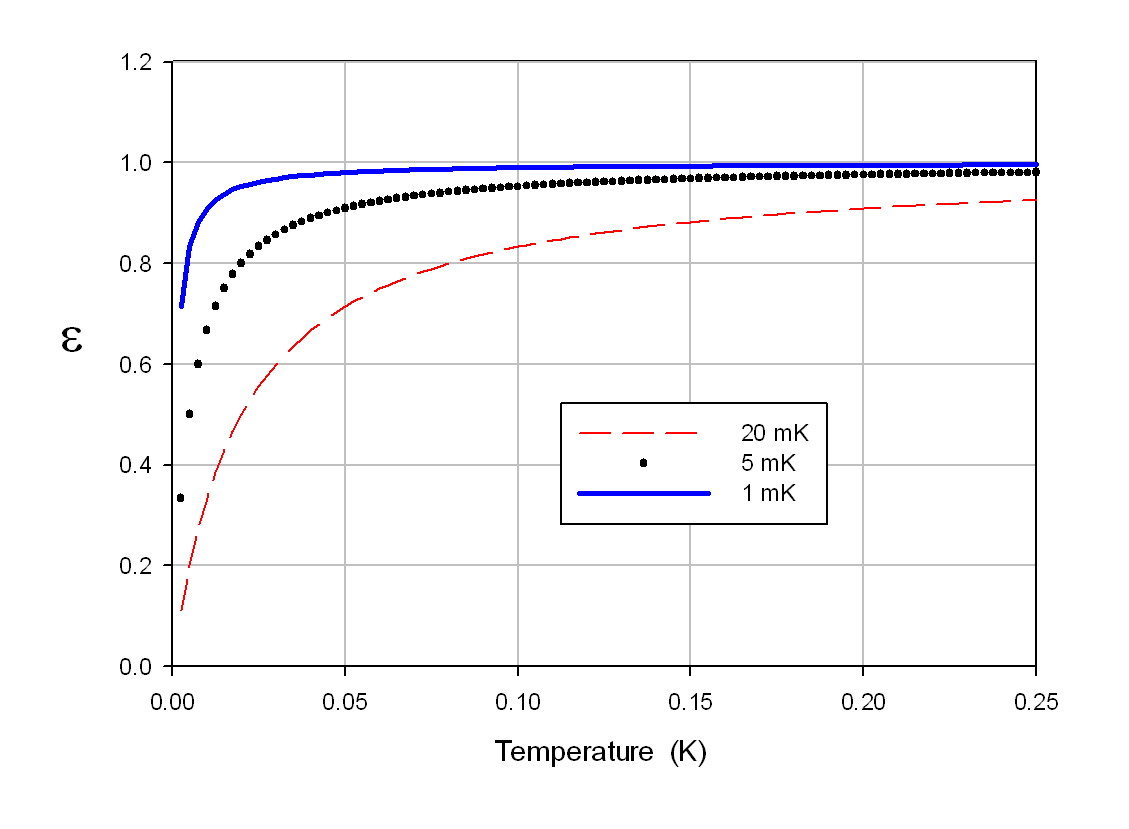}
\caption{Escape rate enhancement $\protect\varepsilon $
as a function of bath temperature from Eq.(4) for various values of $%
E_{N}/k_{B}$.}
\label{noisetemp}
\end{center}
\end{figure}

Therefore consider a \textit{fixed} residual noise energy $E_{N}$ which
combines with the thermal energy $E_{T}=k_{B}T_{B}$, where $T_{B}$ is the
bath temperature, to activate escapes from the Josephson potential well.
Then from Eq.(\ref{Kramers2}), 
\begin{equation*}
\frac{\varepsilon }{k_{B}T_{B}}=\frac{1}{k_{B}T_{B}+E_{N}}
\end{equation*}%
Hence, 
\begin{equation}
\varepsilon =\frac{T_{B}}{T_{B}+E_{N}/k_{B}}  \label{epsilonT}
\end{equation}%
note that the noise factor $E_{N}/k_{B}$ has dimensions of temperature. This
expression quantifies the inherent temperature dependence of $\varepsilon $
in terms of both bath temperature and $E_{N}/k_{B}$. \ Clearly $\varepsilon
\rightarrow 1$ as $E_{N}\rightarrow 0$. Fig.\ref{noisetemp} shows $%
\varepsilon (T_{B})$ curves for various values of $E_{N}/k_{B}$.

There are two procedures with which the noise factor $E_{N}/k_{B}$ may be
estimated from experimental data.

\begin{enumerate}
\item From Eq.(\ref{epsilonT}) $\varepsilon $ will equal $1/2$ when the bath
temperature is equal to the noise factor; hence $T_{B}=E_{N}/k_{B}$.
Considering the experimental data for Oelsner et al. \cite{Oelsner} in Fig.%
\ref{epsilon}, $\varepsilon =0.50$ for $T_{B}\approx 20mK$ and so the noise
factor for that experiment must be $E_{N}/k_{B}\approx 20mK$.

\item Even if the bath temperature were to reach zero, the energy from noise
would still be available to enable escapes from the well. Data points for
Oelsner et al shown in Fig.\ref{adjustment} extrapolate to a limiting SCD
peak of $\eta _{P}\approx 0.965$ at $T_{B}=0$. \ It is then only necessary
to examine the simulation characteristic $\eta _{P}$ vs $T$ as shown in Fig.%
\ref{OelsnerSCD} (solid line) to obtain a temperature corresponding such a
peak position.; hence $E_{N}/k_{B}\approx 30mK$.
\end{enumerate}

From Fig.\ref{epsilon} it is apparent that the residual noise in Oelsner et
al. is the lowest of the three experiments.

\section{Summary}

At the lowest temperatures, experiments exhibit a \emph{smooth} transition
away from the predictions of the standard Kramers' escape rate, Eq.(\ref%
{Kramers}). Figure \ref{adjustment} clearly indicates that an \emph{increase}
of the escape rate, specified by $\varepsilon $ in Eq.(\ref{Kramers2}), is
needed to match simulations with experimental data at these\ low
temperatures. The procedure to determine $\varepsilon $ is based on a fast
algorithm \cite{footnote} that simulates a swept bias experiment and yields
the SCD peak for any selected temperature. \ The obvious consistency of this
approach over the three experiments \cite{VossWebb},\cite{Yu},\cite{Oelsner}%
, as indicated in Fig.\ref{epsilon}, strongly supports this classical model.

The enhancement factor $\varepsilon $ has been shown to result from the
combination of two sources of energy that drive escapes from the Josephson
potential well: the sample at temperature $T$ and residual external noise at
an equivalent tenperature $E_{N}/k_{B}$. For the test case of Oelsner et al. 
\cite{Oelsner} the temperature dependence of $\varepsilon $ derived from
this model (Fig.\ref{noisetemp}) is in good agreement with the empirical
temperature dependence (Fig.\ref{epsilon}) obtained from matching
simulations to experimental data.

Deviations of observed SCD peaks from predictions based on standard escape
rates, at low temperatures, have until now been interpreted as a sign of a
crossover from a classical to a macroscopic quantum state of the Josephson
junction. However as shown in Fig.\ref{corrected}, the near perfect
agreement with experiment of predictions based solely on noise enhanced
thermal activation serve as clear evidence that classical activation remains
the mechanism by which a Josephson junction switches to a finite voltage
(running) state.


\begin{thebibliography}{99}
\bibitem{footnote} Using the algorithm, it is possible to simulate switching
current distribution (SCD) peaks based on an expression for the escape rate,
together with just three parameters: the junction critical current $I_{C}$,
the junction capacitance $C_{J}$ (which sets the zero bias plasma frequency $%
f_{J0}$ ), and the bias current sweep rate. \ The simulation performs the
equivalent of 100,000 bias sweeps with a current resolution of $I_{C}/5000$.
The double precision program runs in Lahey F77 Fortran on a Windows laptop.
\ It returns peak positions and peak widths for $100$ temperatures in about $%
2$ seconds.
\end{thebibliography}
\end{document}